\documentclass[a4paper]{jpconf}
\usepackage{graphicx}
\usepackage{wrapfig}
\usepackage{xspace}

\begin{document}
\title{Recent advancements of the experiment to search for $2K$-capture
in $^{124}$Xe using a
Large Low-background Proportional Counter}

\author
{O.D.~Petrenko$^{2}$, I.D.~Fedorets$^{2}$, A.M.~Gangapshev$^{1}$, Y.M.~Gavrilyuk$^{1}$,
V.V.~Kazalov$^{1}$,V.V.~Kuzminov$^{1}$, S.I.~Panasenko$^{2}$, S.S.~Ratkevich$^{2}$
}
\address{$^1$ Baksan Neutrino Observatory INR RAS, Russia}
\address{$^2$ V.N.Karazin Kharkiv National University, Ukraine}

\ead{petrenko1694@gmail.com }

\begin{abstract}
A new method for correcting the charge signal from the large proportional counter is proposed. The correction method makes it possible to take into account the loss of primary ionization electrons in the presence of weakly increasing microimpurities of electronegative gases during long-term measurements.
It allows increasing the efficiency of the selection of useful events and thereby increasing the sensitivity to $2\nu2K$-decay in $^{124}$Xe at the level of $2\times10^{22}$ from over six years of measurements using a large proportional counter. Analysis of the calibration data, taking into account the new method for correcting the energy release in the detector, made it possible to register events from double $K$ photoionization by a single photon.
\end{abstract}

\section{Introduction}

The experimental signature of the capture of two electrons from the inner shell of $^{124}$Xe is the registration of two $K$-shell X-rays and an Auger electron.
Unambiguous selection of such a complex event is possible in measurements using a large low background cylindrical proportional counter (LPC).
At the ``$2K$-CAPTURE'' facility of the DULB-4900 underground laboratory, BNO INR RAS \cite{DULB} has been carrying out a long experiment to search for double beta decay processes in $^{124}$Xe for several years \cite{MiPhi,G2018}.
A 10.37 L LPC made of oxygen-free copper is used. The LPC is filled with a pure sample enriched to 21\%
in $^{124}$Xe at pressures up to 5.8 bar without any addition of quenching or accelerating gas added.

The LPC signals were read out with a charge-sensitive amplifier (CSA) from one end of the anode wire and recorded to the computer via
a 50-MHz, 12-bit
waveform digitizer for later analysis.
The pulse shape fully displayed the information on the spatial distribution of the primary ionization charges projected onto the counter radius.
For more reliably extraction of useful three-point events, in addition to good energy resolution, adequate measures must be taken to determine the energy scale and resolution accurately and monitor their stability over the full data acquisition period, and take into consideration the relevant uncertainties entering the statistical analysis for the $2K$-capture search.

\section{Recovery of energy release for primary ionization in LPC}

\begin{figure}[t]
\centering
\includegraphics[width=0.9\textwidth]{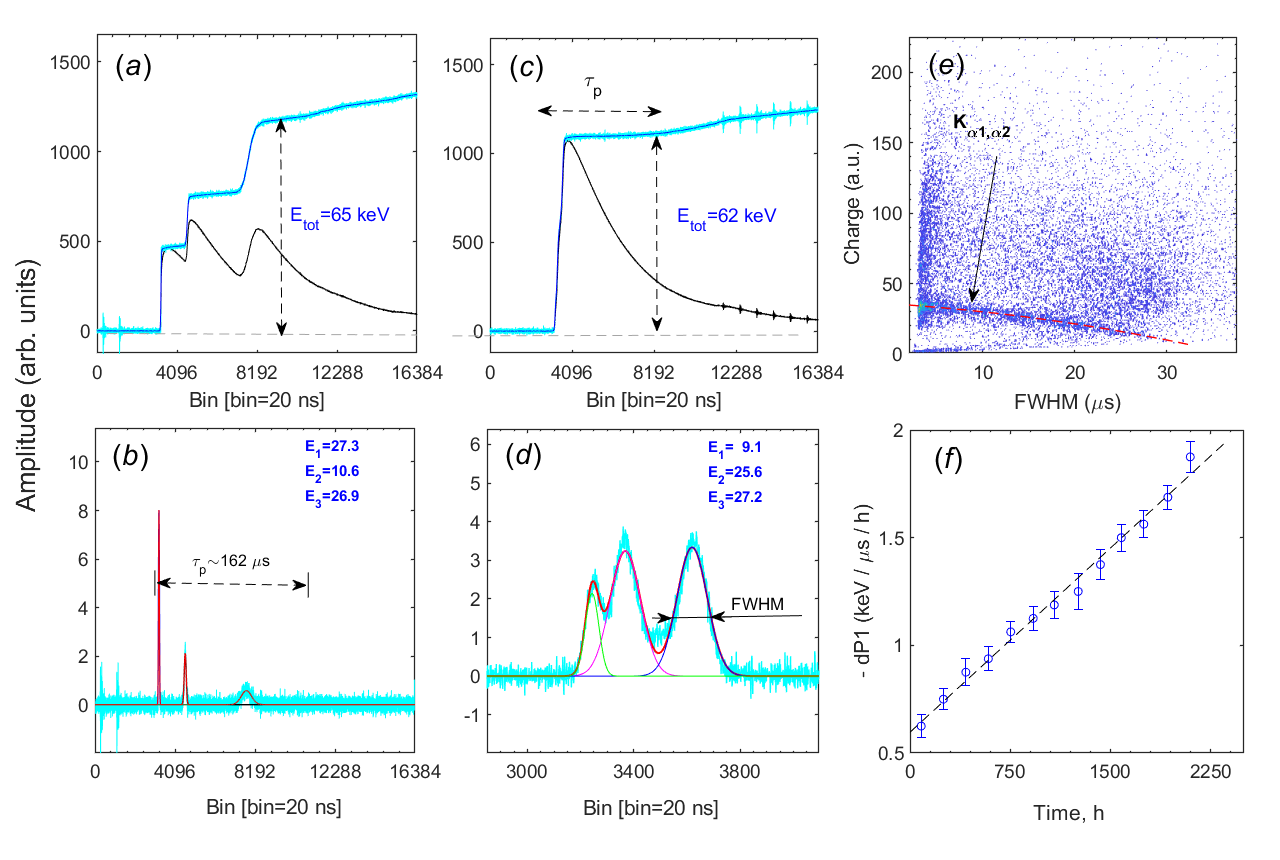} \vspace{-1.75pc}
\caption{{\small\label{pulse3}
Signal from the CSA: $(a)$ and $(c)$ -  the initial charge pulse (dark curve) and the reconstructed signal in the
offline mode (light curves); $(b)$ and $(d)$ - the differentiated charge pulse corresponding to the reconstructed charge signal. $\tau_p$ is the time of electron drift from the cathode to the anode. $(e)$ - bell-shaped signal area (charge ) vs FWHM. $(f)$ - the rate of change degradation factor of the response amplitude from $K$ of X-ray radiation vs the time in the run.
}}
\vspace{-1pc}
\end{figure}

The shape of the electric current pulse is significantly affected by the distribution of the density of primary ionization electrons crossing the boundary of the gas amplification region.
The distribution parameters depend on the drift time of the electrons of the initial point charge to the anode. The primary ionization cluster is blurred during drift due to the radial diffusion of the electron cloud.

Figures \ref{pulse3}$(a)$ and \ref{pulse3}$(c)$ show pulses of three-point events, presumably caused by a $2\nu2K$-capture in $^{124}$Xe.
Black curves represent recorded charge pulses from CSA.
The current pulses of the primary ionization
electrons reconstructed from the digitized charge signals at the boundary of the gas amplification region are shown at Fig.~\ref{pulse3} $(b)$ and $(d)$, the corresponding normalized charge pulses obtained by integrating these current pulses are shown by lighter color curves in Figs.~\ref{pulse3}$(a)$ and \ref{pulse3}$(c)$.
The elapsed time from the onset of the primary pulse to the onset of the first afterpulse $(\tau_p)$ equals the time of electron drift from the cathode to the anode. It presets the duration of the time interval to allocate totally any single event regardless of its primary ionization distribution over the LPC volume. In the case of pure xenon the calculated drift time for the ionization electrons to move from the cathode to the anode is 163 $\mu$s. The integral of the current pulse in the case of a multipoint event for an interval $\tau_p$, starting from the beginning of the pulse, gives the total number of the primary ionization electrons.
A detailed description of the separation method of the electronic and ionic component of the primary ionization cloud from the digitized charge pulse and further extraction of point ionization in a multipoint event is given in Ref.~\cite{PTE}.

A set of bell-shaped curves can describe the resulting current waveform [Fig.~\ref{pulse3}$(b)$ and \ref{pulse3}$(d)$], which allows to determine the charge (energy) released in every point ionization cluster of multipoint event.
The diffusion of primary electrons in the drift region directly affects the distribution of their arrival times in the anode region of high fields. Thus, one can distinguish between events near the cathode surface associated with a long rise time of the charge pulse (wide bell-shaped distributions of the current signal) and events in the internal volume with a shorter rise time (small widths of the current signal).

As an internal detector calibration, we used the photoionization effect by background photons to the $K$-shell of the working gas atoms in the case of $K$ fluorescence.
The two-dimensional distribution of individual amplitudes selected from two-point events with a total energy of 1-200 keV as a function of the
pulse FWHM (rise time) is shown in Fig.~\ref{pulse3}$(e)$.
The locus of photoionization to the Xe-$K$-shell is clearly visible.
This locus makes it possible to determine the degree of degradation of the primary ionization charge as a function of the pulse rise time.
A charge loss correction factor can be obtained by analyzing this relationship for a set of two-week background data collected between two cleanings of gas in the Ti reactor. In this way, the amplitude of each detector response can is corrected.
The rate of change in the degradation coefficient of the detector response to the $K$-series X-rays is shown in Fig.~\ref{pulse3}$(f)$.

\vspace{-0.25pc}
\section{Precision analysis of the LPC response to signals from multipoint ionization}

It is known that, besides the $2K$ capture, a double-$K$-shell photoionization of the atom can create the ``hollow atom'' by absorbing a single photon with the release of two correlated $K$ electrons.
This background events can be used as reference in $2K$-capture experiments, because  background $\gamma$ rays are absorbed there within the fiducial volume of the counter.
In order to select such events, the entire set of calibration measurements from an external $\gamma$-source (see Ref.~\cite{Pet2020}) accumulated
over six years was reanalyzed to test the procedures for processing
background signals in the experiment.
Additionally, the calibration data were analyzed when filling the LPC with krypton to a pressure of 5.5 bar.
During calibration, the counter was irradiated
through the casing wall by photons from
the $^{109}$Cd source with energy 88 keV.
In addition to regular calibration measurements, long-term measurements were performed over several months.

\begin{wrapfigure}[24]{r}{0.52\textwidth} \vspace{-0.75pc}
\hspace{-1.0pc}
\includegraphics[width=0.57\textwidth]{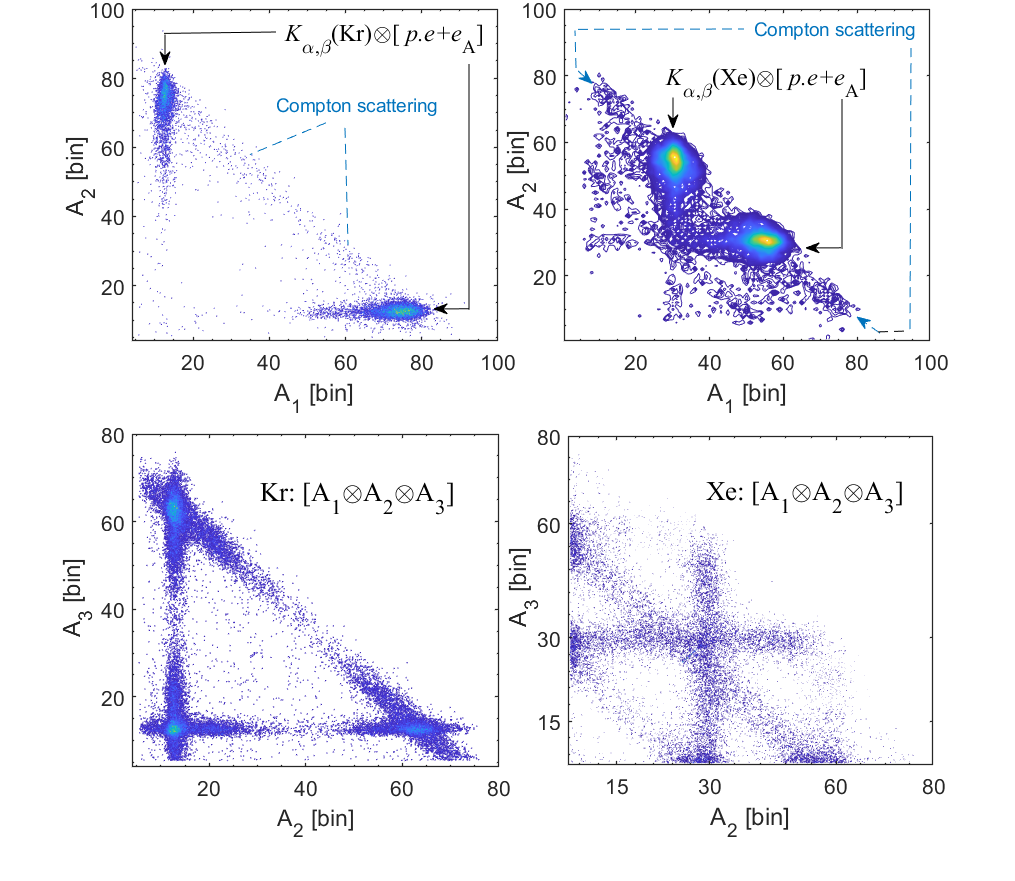} \hspace{-5pc}
\vspace{-2.5pc}
\caption{{\small\label{matr23}
Two-dimensional distributions of the reconstructed amplitudes of energy release in individual point ionization clusters from an external source of 88-keV photons ($^{109}$Cd):
top panels - two-point events and bottom panels - three-point events for krypton and xenon samples, respectively.
}}
\end{wrapfigure}
To confirm our assumptions the characteristics
of two-point events with a total amplitude of 60-95 keV, Fig.~\ref{matr23} (top panels) shows a two-dimensional distribution of these events
depending on the amplitude of the first (A1) and second (A2) subpulses for krypton (left) and xenon (right).
It can be seen from this distribution that intensity maxima are observed for a combination of subpulse amplitudes [$K$-X-ray and photoelectron + Auger electron].
The vertical and horizontal
loci reflect the features of events from incomplete absorption of 88 keV
energy and the shape of the input radiation spectrum.
The oblique diagonal band includes events with total absorption of 88 keV by the photoelectric effect with the production of bremsstrahlung photons and via Compton scattering.
Two oblique diagonal stripes are visible for the xenon target. One band includes events with total absorption of an 88-keV photon by the photoelectric effect with emission $e_A(K)$ and with the formation of bremsstrahlung photons and through Compton scattering. The second band corresponds to an energy of the 58.2 keV and includes events from the peak emission of 29.8 keV from Compton scattering of 88-keV photon by external electrons.
The procedure for fitting a flat substrate into two-dimensional distributions avoids the problem of the Compton background and has the advantage that errors are determined fitting procedure.

The bottom panels of Fig.~\ref{matr23} show two-dimensional distribution of three-point events when convolving in terms of the amplitude of the first subpulse. For a krypton target, pronounced bunches of points are observed at combinations of amplitudes [A2,A3] corresponding to [$K_{\alpha}^h \otimes K_{\alpha}^s$] and [$K_{\alpha}$ $\otimes$ $(\gamma88-2\times K_{\alpha}$)]. For correct explanation of the observed pattern, it should be borne in mind that all three subpulses have identical amplitude spectra. The first region includes events in which the amplitudes of the second and third subpulses correspond to the energy of the $K$-series of a krypton atom.
The invisible first subpulse A1 has an amplitude corresponding to energy 62.8 keV and complements the total energy of the event 88 keV.
The corresponding event selection for the xenon target is shown in the bottom
right panel of the same figure.

Summing the corrected output for each run gives the $I_{1K}(\alpha, \beta)$ value for the normal lines of the $K$-series diagram to obtain the integrated intensity of the dominant single $K$-shell photoionization.
Similarly, sums were carried out on the spectra for three-point events to yield the integrated intensity $I_{2K}(\alpha,\beta)$ for double $K$-shell photoionization.
The total number of registered events was $2.5\times10{^7}$ and $4.2\times10{^7}$ for krypton and xenon, respectively.
The estimate of the total number of double $K$-shell  photoionizations taking
\begin{wrapfigure}[18]{r}{0.4\textwidth} \vspace{-1.5pc} 
\centering
\includegraphics[width=0.4\textwidth]{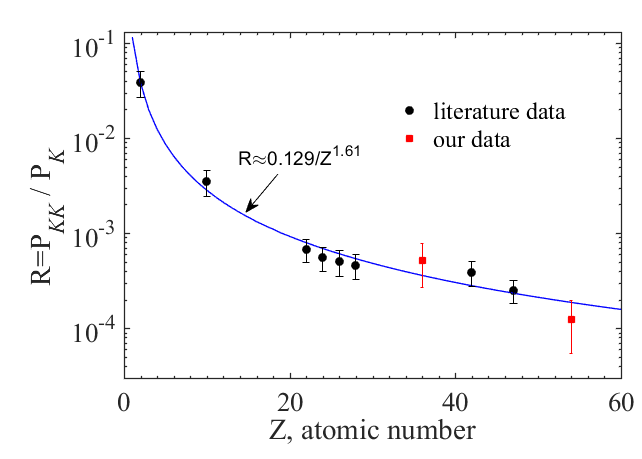} 
\vspace{-2.0pc}
\caption{{\small
\label{ratio_KK}
The ratio of double to single $K$-shell photoionization as function of atomic number. Solid symbols are used for experiments carried out in the asymptotic region for the cross section of double photoionization \cite{Kanter06}.
Solid line is the power-law $1/Z^{1.61}$ fit.}}
\end{wrapfigure}
 into account the fluorescence
yield was
550 and 320 events for krypton and xenon, respectively.
The lower yield for the xenon target can be explained by the fact that the energy of the incident photons was lower than the peak value, in contrast to the krypton target for which $\Delta E=E_\gamma-Q_{2K}$ located in the asymptotic region.

Thus, due to the data obtained, we can estimate the ratio of double and single ionization of $K$-shells by a single photon with an energy of 88 keV for shake-off processes, which was $ 5.2(3)\times10^{-4}$ and $1.2(7)\times10 ^{-4} $ for krypton and xenon, respectively.

Figure \ref{ratio_KK} shows the ratio of the probability of double- to single-photoionization of the $K$-shell as a function of the atomic number.
Black symbols are used for data of experiments carried out in the asymptotic region for the double-$K$-photoionization cross-section \cite{Kanter06}. Red dots show our data obtained.

\vspace{-0.5pc}
\section{Conclusions}
This work demonstrates the testing of a new method for correcting the energy release in multi-cluster ionization in the LPC.
For this, the detector was continuously calibrated for several thousand hours from an external collimated $^{109}$Cd source filled with both krypton and xenon.
Using this technique, we predict an increase in the sensitivity to $2\nu2K$-decay in $^{124}$Xe of $2\times10^{22}$ yr after six years of operation in the current experiment.

\vspace{-0.5pc}

\end{document}